# Physics-Based Compact Modeling of Double-Gate Graphene Field-Effect Transistor Operation

Gennady I. Zebrev, Alexander A. Tselykovskiy, Valentin O. Turin

*Abstract* - An analytic compact model of large-area double-gate graphene field-effect transistor is presented. As parts of the model, the electrostatics of double-gate structure is described and a unified phenomenological approach for modeling of the two drain current saturation modes is proposed.

## I. INTRODUCTION

Graphene field-effect transistors (GFETs), though widely presented in experimental [1, 2, 3] and theoretical studies [4, 5, 6], still require a comprehensive theoretical examination. Typically these transistors are double-gate with a thick back oxide and thin top oxide. The back gate in such structures allows adjusting conductivity type of the channel and controlling the position of current minimum point. In this report we present the model of the graphene double-gate field-effect transistor and use this model for calculation of transistor DC characteristics. The model is based on analytical solution of the current continuity equation in a diffusion-drift approximation [7].

## II. DG GFET ELECTROSTATICS

The energy band diagram of graphene field-effect transistor is shown in Fig. 1.

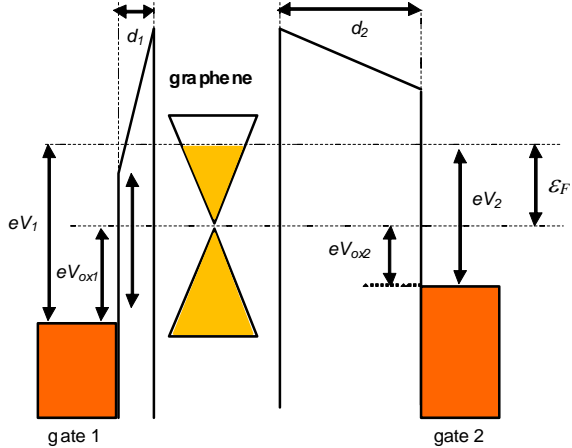

Fig. 1. Double-gate structure energy band diagram

Both gates are positively biased with respect grounded

G.I. Zebrev and A.A. Tselykovskiy are with Department of Micro- and Nanoelectronics, National Research Nuclear University MEPHI, 115409, Kashirskoe sh., 31, Moscow, Russia, E-mail: gizebrev@mephi.ru

V.O. Turin is with the TCAD Laboratory of Orel State Technical University, Orel, Russia

graphene sheet. Charge neutrality condition is given by

$$\varepsilon_1 \varepsilon_0 \frac{eV_1 - \varepsilon_F}{d_1} + \varepsilon_2 \varepsilon_0 \frac{eV_2 - \varepsilon_F}{d_2} = e^2 n_S(\varepsilon_F) + C_{it} \varepsilon_F \quad (1)$$

or, the same, graphene charge density $en_S$ is expressed as

$$e^2 n_S = e(C_1 V_1 + C_2 V_2) - \varepsilon_F(C_1 + C_2 + C_{it}) =$$
$$= (C_1 + C_2)\left( eV_{Geff}(V_1, V_2) - \left(1 + \frac{C_{it}}{C_1 + C_2}\right) \varepsilon_F(V_1, V_2) \right), \quad (2)$$

where $V_{Geff} = (C_1 V_1 + C_2 V_2)/(C_1 + C_2)$ is the effective gate voltage, $V_{1(2)}$ is the top (back) gate voltage, $V_{1(2)} = V_{G1(2)} - V_{NP1(2)}$, $V_{NP1(2)}$ are charge neutrality biases, $\varepsilon_F$ is Fermi energy in graphene, $C_{1(2)}$ is the sheet capacitance of top (back) oxide, $C_{it}$ is the interface traps capacitance assumed here to be energy independent. We found an explicit dependence of the Fermi energy as function of the both gate voltages

$$\varepsilon_F(V_1, V_2) = \left( m^2 \varepsilon_{ad}^2 + 2\varepsilon_{ad} eV_{Geff} \right)^{1/2} - m\varepsilon_{ad}, \quad (3)$$

$$m = 1 + \frac{C_{it}}{C_1 + C_2}, \quad \varepsilon_{ad} = \frac{\pi \hbar^2 v_0^2 (C_1 + C_2)}{2e^2}, \quad (4)$$

which represent simple generalizations of the parameters defined for a single gate case (if one oxide capacitance much larger than another, all equations transform into a single gate form)[7]. The characteristic energy $\varepsilon_{ad}$ is nothing but the full electrostatic energy stored in both gate capacitors per one carrier in graphene, which turns out to be gate voltage independent for zero-gap material.

Equivalent circuit for double-gate GFET is shown in Fig. 2.

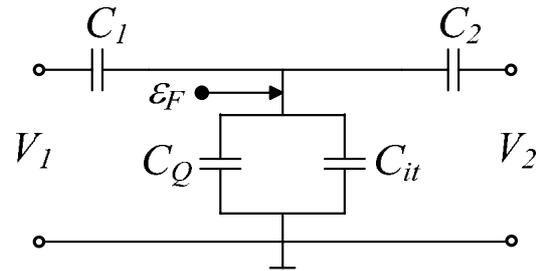

Fig. 2. Equivalent circuit of double-gate GFET structure

The capacitance of the gate 1 (2) per unit area at grounded gate 2 (1) is given by

$$C_{G1(2)} = \left( \frac{1}{C_{1(2)}} + \frac{1}{C_Q + C_{it} + C_{2(1)}} \right)^{-1}, \quad (5)$$

where $C_Q$ is the quantum capacitance, $C_{it}$ is the interface trap capacitance.

We calculate the channel capacitance with respect to a single gate at grounded other gate using Eqs.2,3 and 4

$$C_{CH1(2)} \equiv \left(\frac{e\partial n_S}{\partial V_{1(2)}}\right)_{V_{2(1)}} = \frac{C_{1(2)} C_Q}{C_Q + C_1 + C_2 + C_{it}} \quad (6)$$

The gate and the channel capacitances are interrelated in graphene gated structures through exact relation

$$\frac{C_{G1(2)}}{C_{CH1(2)}} = 1 + \frac{C_{it} + C_{2(1)}}{C_Q}. \quad (7)$$

Notice that the capacitance of the grounded gate is in parallel connection with the interface trap capacitance. Following Ref. [7] we are about to obtain an explicit analytical solution of continuity equation for channel current density. Total drain current $J_S = J_{DR} + J_{DIFF} = (1+\kappa)J_{DR}$ should be conserved along the channel

$$\frac{dJ_S}{dy} = 0 \Leftrightarrow \frac{d}{dy}(n_S E) = 0 \quad (8)$$

that yields an equation for electric field distribution along the channel[*]

$$\frac{dE}{dy} = \left(\frac{e\,dn_S}{n_S d\zeta}\right)\left(-\frac{d\zeta}{ed\varphi}\right)\left(\frac{d\varphi}{dy}\right)^2 = \frac{\kappa e}{\varepsilon_D} E^2. \quad (9)$$

where the "diffusion energy" $\varepsilon_D = n_S / (dn_S / d\varepsilon_F)$ and diffusion-to-drift currents ratio $\kappa = J_{DIFF} / J_{DR}$ are assumed to be functions of only the gate voltage rather than the drain-source bias and position along the channel.

To properly derive explicit expression for control parameter $\kappa$ we have to use the electric neutrality condition along the channel length in gradual channel approximation which is assumed to be valid even under non-equilibrium condition $V_{DS} > 0$. Acting similarly as in a single-gated structure one can get

$$\kappa = -\left(\frac{\partial \zeta}{e\partial \varphi}\right)_{V_{Geff}} = \frac{(\partial V_{Geff}/\partial \varphi)_\zeta}{e(\partial V_{Geff}/\partial \zeta)_\varphi} = \frac{C_1 + C_2}{C_Q + C_{it}} \quad (10)$$

This dimensionless parameter $\kappa$ is assumed to be constant along the channel for a given electric biases and expressed via the ratio of characteristic capacitances. For ideal graphene channel with low interface trap density the $\kappa$-parameter is a function of only $\varepsilon_{ad}$ and the Fermi energy

$$\kappa(C_{it} = 0) = \frac{C_1 + C_2}{C_Q} = \frac{\varepsilon_{ad}}{\varepsilon_F}. \quad (11)$$

For a high-doped regime (large $C_Q$) and/or thick gate oxide (low $C_{ox}$) when $C_Q \gg C_{ox}$ we have $\kappa \ll 1$ and the drift current component dominates the diffusion one and vice versa.

---

[*] Notice a typo in Eq.65 of Ref.[7]: lost a power 2 in an intermediate relation

Straightforward solution of ordinary differential Eq. 9 yields

$$E(y) = \frac{E(0)}{1 - \frac{\kappa e^2 E(0)}{\varepsilon_D} y}, \quad (12)$$

where $E(0)$ is electric field near the source, which should be determined from the condition imposed by a fixed electrochemical potential difference between drain and source $V_{DS}$, playing a role of boundary condition

$$V_{DS} = (1+\kappa) \int_0^L E(y) dy, \quad (13)$$

where $L$ is the channel length. Using Eqs. (12) and (13) one obtains an expressions for $E(0)$ and electric field distribution along the channel

$$E(0) = \frac{\varepsilon_D/e}{\kappa L}\left(1 - \exp\left(-\frac{\kappa}{1+\kappa}\frac{eV_{DS}}{\varepsilon_D}\right)\right); \quad (14)$$

$$E(y) = \frac{\frac{\varepsilon_D/e}{\kappa L}\left(1 - \exp\left(-\frac{\kappa}{1+\kappa}\frac{eV_D}{\varepsilon_D}\right)\right)}{1 - \frac{y}{L}\left(1 - \exp\left(-\frac{\kappa}{1+\kappa}\frac{eV_D}{\varepsilon_D}\right)\right)}. \quad (15)$$

### III. CURRENT-VOLTAGE CHARACTERISTICS

According general rules the total current at constant temperature can be written as gradient of the electrochemical potential taken in the vicinity of the source

$$I_D = eW \mu_0 n_S(0)(1+\kappa)E(0) =$$
$$= e\frac{W}{L} D_0 n_S(0) \frac{1+\kappa}{\kappa}\left(1 - \exp\left(-\frac{\kappa}{1+\kappa}\frac{eV_{DS}}{\varepsilon_D}\right)\right), \quad (16)$$

where $\mu_0$ is the low-field carrier's mobility, $W$ is the channel width, and the Einstein relation $D_0 = \mu_0 \varepsilon_D / e$ is employed. Defining a saturation current drain voltage as

$$V_{DSAT} = 2\frac{1+\kappa}{\kappa}\frac{\varepsilon_D}{e} = \frac{1+\kappa}{\kappa}\frac{\varepsilon_F}{e}, \quad (17)$$

we obtain

$$V_{DSAT}(V_1, V_2) = V_{Geff} + en_S(V_1, V_2)/(C_1 + C_2). \quad (18)$$

Integration of Eq.15 yields the explicit relationship for distribution of the electric, chemical and electrochemical $\mu = \zeta - e\varphi$ potential

$$\mu(y) - \mu(0) = \frac{eV_{DSAT}}{2}\log\left[1 - \frac{y}{L}\left[1 - \exp\left(-\frac{2V_{DS}}{V_{DSAT}}\right)\right]\right], \quad (19)$$

where $\mu(0)$ is the electrochemical potential nearby the source controlled by the gate-source bias $V_{GS}$. For any gate voltage $V_{GS}$ (and corresponding $\kappa(V_G)$) the full drop of electrochemical potential $\mu$ on the channel length is fixed by the source-drain bias $\mu(L) = \mu(0) - eV_{DS}$. General relation for drain current (Eq.16) can be rewritten using low-field conductance given by

$$g_{D0} = (W/L)e\mu_0 n_{S0} \equiv (W/L)\sigma_0$$ ($n_{S0}$ is the carrier density nearby the source) as

$$I_D = \frac{1}{2} g_{D0} V_{DSAT} \left(1 - \exp\left(-2\frac{V_{DS}}{V_{DSAT}}\right)\right). \quad (20)$$

## IV. Two Current Saturation Modes

The field-effect transistor is fundamentally non-linear device working at large biases generally on all electrodes. The saturation of the channel current in the FETs at high source-drain electric field has two-fold origin, namely, (i) the current blocking due to carrier density depletion near the drain, and (ii) the carrier velocity saturation due to optical phonon emission. The saturation current for pinch-off case arises due to saturation of lateral electric field near the source. Using the Einstein relation in a form $D_0 C_Q = \sigma_0$ the pinch-off saturation current in Eq. 16 may be represented in an alternative form $I_{DSAT} = W e n_{S0} v_S$, where the characteristic velocity is defined as

$$v_S = \frac{\mu_0 V_{DSAT}}{2L}. \quad (21)$$

The current saturation for short-channel FETs (typically $L \leq 0.5$ μm) is bound to the velocity saturation due to scattering on optical phonons [8]. The channel current saturates due to velocity saturation at $I_{DSAT} = W e n_{S0} v_{opt}$. Note, that for the diffusive channels the saturation velocity $v_{opt}$ is a maximum velocity of dissipative motion, which is in any case less than the speed $v_0$ of ballistic carriers in graphene. One can introduce the dimensionless parameter discriminating the two types of current saturation in FET [9]

$$a = \frac{v_S}{v_{opt}} = \frac{\mu_0 V_{DSAT}}{2 v_{opt} L} = \frac{V_{DSAT}}{V_{D0}}, \quad (22)$$

where a new characteristic drain voltage is defined

$$V_{D0} \equiv \frac{2 v_{opt} L}{\mu_0}, \quad (23)$$

$$V_{D0} \cong 10 \left(\frac{2 v_{opt}}{10^8 cm/s}\right)\left(\frac{L}{1 \mu m}\right)\left(\frac{10^3 cm^2/Vs}{\mu_0}\right) \text{ V}. \quad (24)$$

Thereby the drain current can be rewritten in a unified manner for both cases

$$I_D = W e n_{S0} v_{SAT} \left(1 - \exp\left[-\frac{\mu_0 V_{DS}}{v_{SAT} L}\right]\right) \quad (25)$$

where $v_{SAT} = \min\{v_{opt}, v_S\}$. A reasonable analytical interpolation can be used:

$$v_{SAT} = v_{opt} \tanh \frac{v_S}{v_{opt}} = v_{opt} \tanh\left(\frac{\mu_0 V_{DSAT}}{2 v_{opt} L}\right), \quad (26)$$

which provides convenient analytical description of crossover between two modes of saturation.

Note, that empirical relationships for high-field drift velocity

$$v_{DR}(E) = \frac{\mu_0 E}{\left(1 + (\mu_0 E / v_{SAT})^n\right)^{1/n}} \equiv \frac{\mu_0 E}{\left(1 + (E / E_{SAT})^n\right)^{1/n}} \quad (27)$$

originating from the early work of Thornber [10] and traditionally used in CMOS compact modeling [11] also is nothing but empirical interpolation having besides a significant shortage. This equation does not provide fast saturation and yields only $v_{SAT}/2^{1/n}$ at $E = v_{SAT}/\mu_0$. To remove this shortage for best fitting with experiments a joint interpolation is typically used in CMOS design practice with $E_{SAT} = 2 v_{SAT}/\mu_0$ and artificial fitting to obey a formal condition $v(E_{SAT}) = v_{SAT}$. A use of analytic interpolation Eq.26 allows to get rid of piecewise description and senseless fitting parameter $n$.

Description of the two saturation modes can be combined by the unified expression for the drain current as function of the drain-source voltage

$$I_D = \frac{1}{2} g_{D0} V_S \left(1 - \exp\left(-2\frac{V_{DS}}{V_S}\right)\right), \quad (28)$$

where generalized saturation source-drain voltage

$$V_S = V_{D0} \tanh \frac{V_{DSAT}}{V_{D0}}. \quad (29)$$

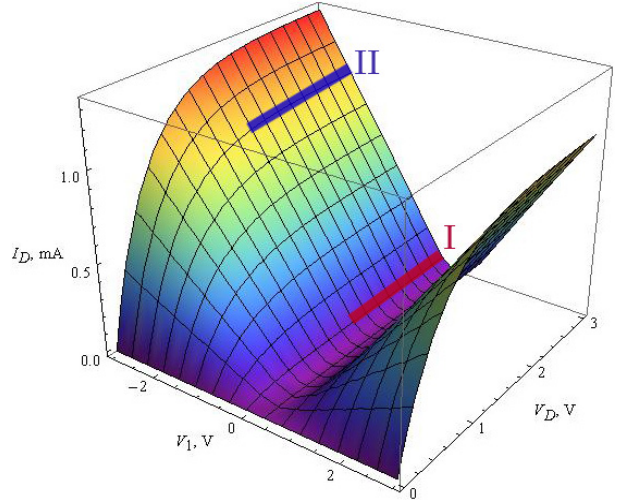

Fig. 3. Simulated I-V characteristics of GFET (a) as function of front gate $V_1$ and drain voltages $V_D$ calculated at $V_2 = -30$ V, $\mu_0 = 1000$ cm$^2$/(V×s), $\varepsilon_1 = 16$, $d_1 = 15$ nm, $\varepsilon_2 = 4$, $d_2 = 300$ nm, $W = 1$ μm, $L = 1$ μm, $C_{it} = 0$, $v_{opt} = 5 \times 10^6$ cm/s; (I) $a = 0.6$ (electrostatic pinch-off), (II) $a = 4.7$ (velocity saturation).

At small $V_{DS}$ ($V_{DS} << V_S$) the drain current is determined by only the low-field conductance $g_{D0}$. The dimensionless parameter $a$ discriminates the two types of current saturation in the FETs at large $V_{DS}$. When $a << 1$ (long channel and thin gate insulators, low carrier density and mobility) the electrostatic pinch-off prevails ("square law"), and if $a \gg 1$ the carrier velocity saturation determines the saturation current of FETs

$$I_{SAT} \cong W e n_S v_{opt} \quad (30)$$

The drain current saturation mode depends on geometrical and transport parameters of the transistor ($V_{D0}$) as well as on the electric operation mode since $V_{DSAT}$ is a function of the gate voltages. Figs. 3-4 show calculated current-voltage characteristics exhibiting different modes of drain current saturation.

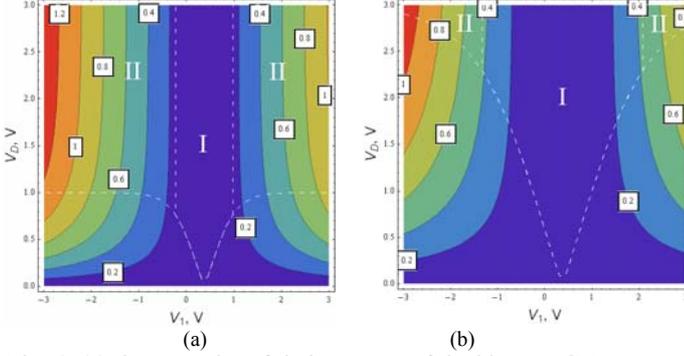

Fig. 4. (a) Contour plot of drain current of double-gate GFET as function of front gate and drain voltages; $V_2 = -30$ V, $\mu_0 = 5000$ cm$^2$/(V×s), $\varepsilon_1 = 16$, $d_1 = 15$ nm, $\varepsilon_2 = 4$, $d_2 = 300$ nm, $C_{it} = 0$, $v_{opt} = 5 \times 10^7$ cm/s, $W = 1$ μm, (a) $L = 0.5$ μm ($V_{D0} = 1$ V), (b) $L = 1.5$ μm ($V_{D0} = 3$ V). Dashed lines separate different drain current modes: (I) the electrostatic pinch-off, (II) velocity saturation; (III) no saturation. The numbers in the white rectangles are the drain current values in mA.

## VI. INTRINSIC OUTPUT CONDUCTANCE AND TRANSCONDUCTANCE OF DOUBLE-GATE GFETs

Ignoring many complications one can conclude that current-voltage characteristics with saturation may easily parameterized by the two parameters: the output conductance and the saturation voltage. The drain conductance as function of the node biases (closely connected with low-field conductance $g_{D0}$) can be calculated as a partial derivative of drain current with a fixed $V_{GS}$

$$g_D = \left(\frac{\partial I_D}{\partial V_{DS}}\right)_{V_{GS}} = g_{D0} \exp\left(-\frac{2V_{DS}}{V_{S0}}\right). \quad (31)$$

One of the most important small-signal parameter for high-frequency performance prediction is the intrinsic gate transconductance $g_m$. Transconductance depends generally on microscopic mobility slightly varying with the gate voltage the underlying mechanism and quantitative description of that has not been yet developed in details. Omitting here this point the microscopic mobility will be considered as to be independent on the gate bias in this report. Exact view of relation of the intrinsic transconductance for arbitrary value of the parameter $a$ depends on the choice of approximation for current and has awkward form. We will use here a convenient approximation for both gates

$$g_{m1(2)} \cong \frac{W}{2L} \mu_0 C_{CH1(2)} V_{S0} \left(1 - \exp\left(-\frac{2V_{DS}}{V_{S0}}\right)\right). \quad (32)$$

The transconductance $g_m$ increases linearly with $V_{DS}$ up to saturation on a maximum level

$$g_{m1(2)}^{(max)} \cong \frac{W}{2L} \mu_0 C_{CH1(2)} V_{S0} =$$
$$= \begin{cases} \frac{W}{2L} \mu_0 C_{CH1(2)} V_{DSAT} = W C_{CH1(2)} v_S, & V_{D0} > V_{DSAT} \\ \frac{W}{2L} \mu_0 C_{CH1(2)} V_{D0} = W C_{CH1(2)} v_{opt}, & V_{DSAT} > V_{D0}. \end{cases} \quad (33)$$

Access and parasitic contact resistances can significantly degrade extrinsic performance characteristics of GFETs. We ignore yet the charge multiplication effects with characteristic super-linear dependence on drain voltage in this report. These effects occurs typically at high $V_{DS}$ and low charge densities in graphene when the channel driven electric fields nearby the drain are maximum and validity of semi-classical diffusion-drift approximation is failed.

## REFERENCES

[1] I. Meric et al., 2008, "Current saturation in zero-bandgap, top-gated graphene field- effect transistors," Nature Nanotech. 3, 654–659.
[2] I. Meric, C. Dean, A. F. Young, J. Hone, P. Kim, and K. L. Shepard, "Graphene field-effect transistors based on boron nitride gate dielectrics," International Electron Devices Meeting, 2010, pp. 23.2.1-23.2.4.
[3] S.-J. Han, Z. Chen, A.A. Bol, and Y. Sun, "Channel-Length-Dependent Transport Behaviors of Graphene Field-Effect Transistors," *IEEE Electron Device Lett.*, vol. 32, no. 6, pp. 812–814, June 2011.
[4] S. Thiele, J. A. Schaefer, & F. Schwierz, "Modeling of graphene metal–oxide–semiconductor field-effect transistors with gapless large-area graphene channels," *J. Appl. Phys.* **107**, 094505 (2010).
[5] S. Thiele and F. Schwierz, "Modeling of the steady state characteristics of large-area graphene field-effect transistors," *J. Appl. Phys.* 110, 034506 (2011); doi:10.1063/1.3606583.
[6] J.G. Champlain, "A first principles theoretical examination of graphene-based field effect transistors," *J. Appl. Phys.* **109**, 084515 (2011).
[7] G. I. Zebrev, "Graphene Field Effect Transistors: Diffusion-Drift Theory", in "*Physics and Applications of Graphene – Theory,*" Ed. by S. Mikhailov, Intech, 2011.
[8] I. Meric et al. "RF performance of top-gated, zero-bandgap graphene field-effect transistors," IEDM, 2008
[9] G.I. Zebrev, "Current-voltage characteristics of a metal-oxide-semiconductor transistor calculated allowing for the dependence of the mobility on a longitudinal electric field," Fiz. Tekh. Poluprovodn. (Sov. Phys. Semicond.),vol. 26, no.1, (1992).
[10] K.K. Thornber, "Relation of drift velocity to low-field and high-field saturation velocity," J. Appl. Phys. 1980, 51, 2127.
[11] Y. Cheng, C. Hu, "MOSFET Modeling & BSIM3 User's Guide," Kluwer Academic Publishers, 2002.